\documentclass[ %
 reprint,
superscriptaddress,
 amsmath,amssymb,
 aps,
prb,
]{revtex4-1}

\usepackage{graphicx}
\usepackage{dcolumn}
\usepackage{bm}
\usepackage{subfigure}
\usepackage{epstopdf}
\usepackage{lettrine}

\begin{document}

\preprint{APS/123-QED}

\title{Incipient singlet-triplet states in a hybrid mesoscopic system}

\author{Chengyu Yan}
 \email{uceeya3@ucl.ac.uk}
 \affiliation{%
 London Centre for Nanotechnology, 17-19 Gordon Street, London WC1H 0AH, United Kingdom\\
 }%
 \affiliation{
  Department of Electronic and Electrical Engineering, University College London, Torrington Place, London WC1E 7JE, United Kingdom
 }%
 \author{Sanjeev Kumar}
 \affiliation{%
 London Centre for Nanotechnology, 17-19 Gordon Street, London WC1H 0AH, United Kingdom\\
 }%
 \affiliation{
  Department of Electronic and Electrical Engineering, University College London, Torrington Place, London WC1E 7JE, United Kingdom
 }%
\author{Michael Pepper}
\affiliation{%
 London Centre for Nanotechnology, 17-19 Gordon Street, London WC1H 0AH, United Kingdom\\
 }%
 \affiliation{
  Department of Electronic and Electrical Engineering, University College London, Torrington Place, London WC1E 7JE, United Kingdom
 }%
\author{Patrick See}
\affiliation{%
 National Physical Laboratory, Hampton Road, Teddington, Middlesex TW11 0LW, United Kingdom\\
}%
\author{Ian Farrer}
\affiliation{%
	Cavendish Laboratory, J.J. Thomson Avenue, Cambridge CB3 OHE, United Kingdom\\
}%
\author{David Ritchie}
\affiliation{%
 Cavendish Laboratory, J.J. Thomson Avenue, Cambridge CB3 OHE, United Kingdom\\
}%
\author{Jonathan Griffiths}
\affiliation{%
 Cavendish Laboratory, J.J. Thomson Avenue, Cambridge CB3 OHE, United Kingdom\\
}%
\author{Geraint Jones}
\affiliation{%
 Cavendish Laboratory, J.J. Thomson Avenue, Cambridge CB3 OHE, United Kingdom\\
}%

\date{\today}
             
\begin{abstract}
	
In the present work we provide an easily accessible way to achieve the singlet-triplet Kondo effect in a hybrid system consisting of a quantum point contact (QPC) coupled to an electronic cavity. We show that by activating the coupling between the QPC and cavity, a zero-bias anomaly occurs in a low conductance regime, a coexistence of zero-bias and finite-bias anomaly (FBA) in a medium conductance regime, and a FBA-only anomaly in a high conductance regime. The latter two observations are due to the singlet-triplet Kondo effect. 
 
\end{abstract}

\maketitle

\section{Introduction}
The ordered phase of electrons, where electrons localize themselves in a particular period or lattice, such as the skyrmion lattice or the Wigner lattice, has attracted considerable interest\cite{GST05,KLM14,DLP16,JDL18}. The existence of ordered phase in a two-dimensional (2D) system has been successfully probed by geometric resonance in magneto-oscillations\cite{GST05,KLM14,DLP16,JDL18}. Despite observations of the incipient Wigner lattice in one-dimensional (1D) system by means of conductance measurements\cite{HTP09, KTS14}, it is still challenging to monitor the ordered phase in a 1D system. This may be due to a merger with the 2D Fermi sea as the 1D electrons leave the 1D regime, thus losing their ordered phase. The first step towards realizing an ordered 1D phase could be forming a chain of localized electrons, and in this regard the multi-impurity Kondo effect appears to be an useful tool to visualize the formation of localized electrons. The multi-impurity Kondo effect arises from coherent spin-flip scattering between the conduction and multiple localized electrons\cite{KONDO64,PRM06, RFB08}.  For the odd-numbered Kondo effect, screening of the unpaired spin-1/2 localized electron gives rise to a zero-bias anomaly (ZBA) in the differential conductance. For the even-numbered Kondo effect, the spin-configuration $|S,m>$ of the localized electrons is vital ($S$ is the total spin and $m$ is the spin projection). In the singlet regime ($|0,0>$), a finite-bias anomaly (FBA) occurs due to singlet-triplet transition while no ZBA is allowed\cite{PRM06, RFB08}; on the other hand, both FBA and ZBA (a partially screened spin-1 Kondo effect in this particular case\cite{SDE00}) can be observed in the triplet regime\cite{RFB08} ($|1,1>$, $|1,0>$ and $|1,-1>$).

The occurrence of FBA in the singlet regime and coexistence of FBA and ZBA in the triplet regime have been observed in quantum dots (QD)\cite{PRM06, RFB08}. On the other hand, less progress has been achieved in quantum point contacts (QPC) owing to difficulties in probing the spin configuration of the localized electrons. Some recent work in QPCs illustrate an abnormal splitting of ZBA \cite{ILK13,BMF14} in a narrow conductance window around $0.8 \times \frac{2e^2}{h}$. However, the coexistence of FBA and ZBA was absent in such cases.  Also, it has been shown the non-Kondo disorder within the 1D channel can also result in the splitting of ZBA\cite{SSD09}. Therefore, the understanding of the double-impurity Kondo effect (or singlet-triplet Kondo effect) in QPCs is far from complete.

\begin{figure}
   
	\includegraphics[height=3.6in,width=3.2in]{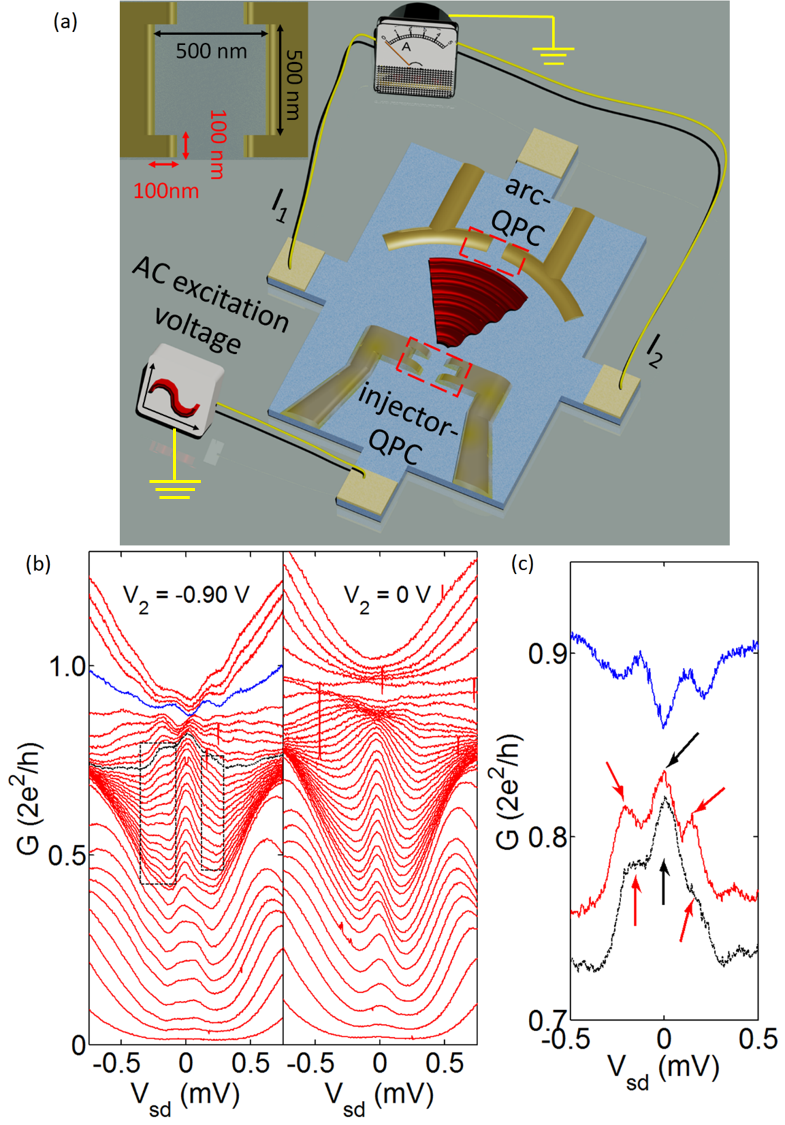} 

	\caption{Setup of the experiment and main result. (a) Schematic of the setup of nexperiment. The square gold pads at the end of the mesa are ohmic contacts; the opening angle of the arc-shaped split gate is 45$^\circ$ and the radius is 2.0 $\mu$m, both the length and width of QPC embedded in the arc (hereafter referred as arc-QPC) is 200 nm; The length (width) of the injector-QPC is 700 nm (500 nm). The shining red patten represents the Friedel oscillations. The current meter measures I$_1$ + I$_2$. The inset shows a zoom-in of the injector-QPC in sample A-C, whereas a conventional rectangular QPC is used in sample D. (b) It is seen that with cavity switched on ($V_2$ = -0.90 V), FBAs (highlighted by the black dashed box) are observed along with the ZBA; on the other hand, only the ZBA is present with cavity switched off (($V_2$ = 0 V). It should be noted that in order to illustrate the main features, traces overlapping together around 0.85$\times \frac{2e^2}{h}$ are selectively plotted. (c) A zoom-in of representative traces of the first panel in (b); the arrows highlight the ZBA (black) and FBA (red). }
	\label{fig:1}
\end{figure}

 \begin{figure}

	\includegraphics[height=3.2in,width=3.6in]{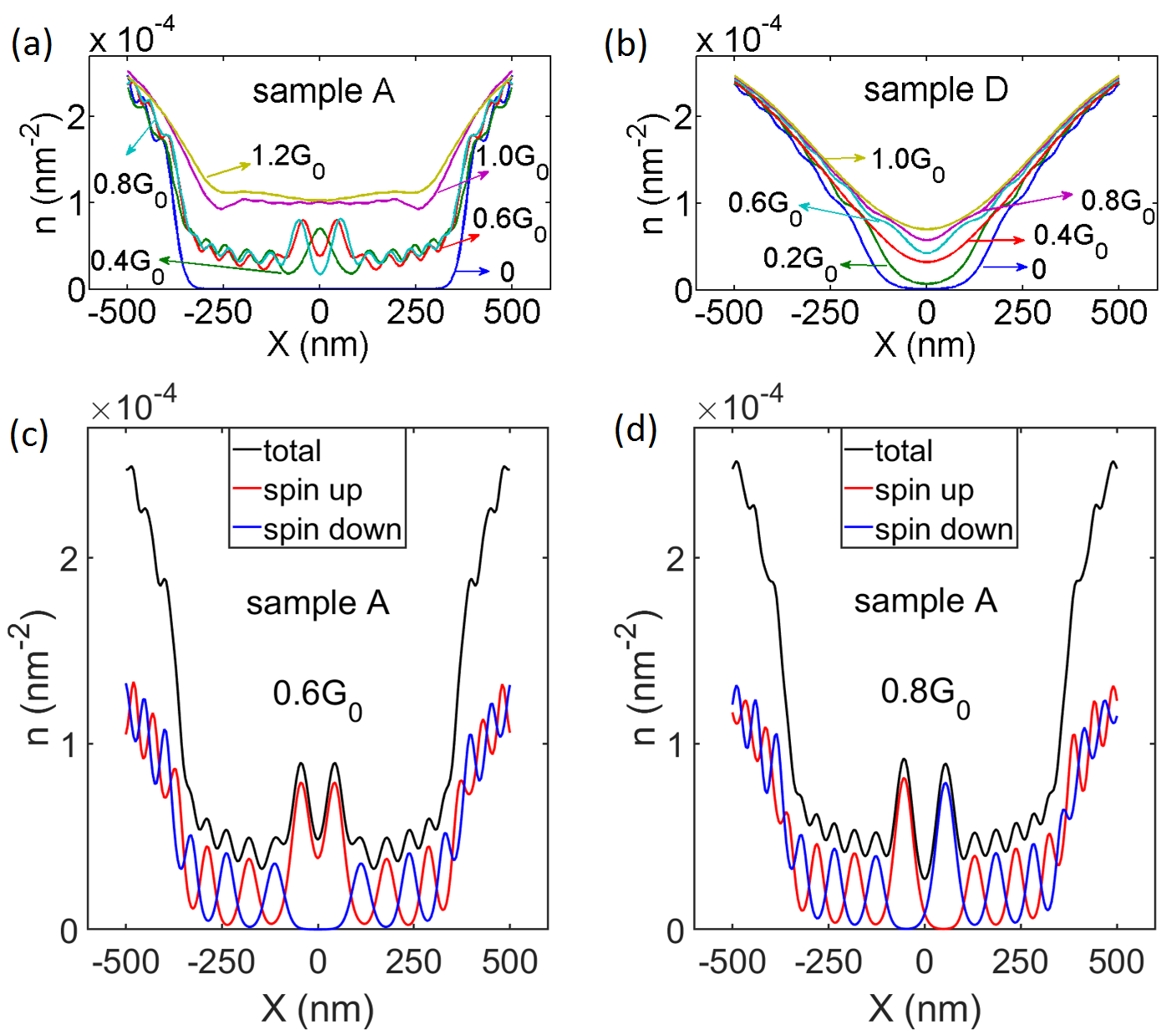} 
	
	\caption{Simulated electron density along the current flow direction. (a) and (b) Simulated electron at different injector conductance for sample A and D, respectively. (c) and (d) Spin configuration in sample A with V$_1$ set to 0.6G$_0$ and 0.8G$_0$, respectively.
	} 
	\label{fig:2}
\end{figure} 

Here we provide an easily accessible route to  realize the singlet-triplet Kondo effect in a hybrid system consisting of a QPC coupled to an electronic cavity. The cavity refocuses the injected electron back to the QPC\cite{ROZ15} and thus tunes the effective electron density within the QPC  without effectively changing the electrostatic potential. We show in this system the coexistence of ZBA and FBA in the moderate conductance regime in addition to the  splitting of ZBA in high conductance regime. Our results also indicate that the occurrence of the 0.7 conductance anomaly, which has been a subject of continuous debate, is not correlated with ZBA or FBA. 

\section{Experiment}
The hybrid devices were fabricated on a high mobility two-dimensional electron gas (2DEG) formed at the interface of GaAs/Al$_{0.33}$Ga$_{0.67}$As heterostructure. The metallic gates are deposited on the surface which is 90 nm away from the 2DEG. The electron density (mobility) measured at 1.5 K was 1.80$\times$10$^{11}$cm$^{-2}$ (2.1$\times$10$^6$cm$^2$V$^{-1}$s$^{-1}$). All the measurements were performed with the standard lock-in technique in a cryofree dilution refrigerator with a lattice temperature of 20 mK. 

The samples studied in the present work consist of a pair of arc-shaped gates, with a QPC forming in the center of the arc, and an injector-QPC as shown in Fig.~\ref{fig:1}(a). The injector-QPC has been shaped to have a slot in the center in sample A-C [inset of Fig.~\ref{fig:1}(a)], whereas for sample D, a conventional QPC with rectangular split gates\cite{KTS14} was used for comparison [see the inset of Fig.~\ref{fig:5}(a)]. Previously, it was shown that a weakly bound state can be often formed in a QPC with protrusions in the split gates\cite{SFP08}. The main results are obtained from sample A while sample B and C show similar behaviour (Supplemental Fig.~S4). It has been carefully examined that the injector-QPC did not show a QD-like behaviour [see Supplemental Fig.~S1(d)\cite{note}].

Before we discuss the main results of the present work, it is necessary to clarify that, despite the injector-QPC on sample A-C looks QD-like but it does not behave like a QD. It was suggested that a QD-like QPC may have three distinctive working regime\cite{HBS15}, namely, a QPC-dominant regime, a QPC-QD transition regime (i.e. device inherits characteristic from both QPC and QD), and a QD-dominant regime, according to the profile of the electrostatic potential. In QPC-QD transition regime, a Fabry-Perot type interference should be present on conductance plateaus, whereas in the QD-dominant regime, Coulomb blockade peaks superpose on conductance plateaus\cite{HBS15}. In our experiment, the conductance plateaus are free of oscillations, as shown in Fig.~\ref{fig:5}(b), and therefore suggests our device is not in QPC-QD transition regime or QD-dominant regime.

\section{Result and discussion}
The hybrid system, as shown in Fig.~\ref{fig:1}(a), exhibited interesting behaviour in the presence of source-drain bias [Fig.~\ref{fig:1}(b)]. In this experiment, the conductance of the injector-QPC was incremented slowly by changing the voltage V$_1$ applied to the injector-QPC for a fixed voltage V$_2$ applied across the arc-QPC. An electronic cavity can be created between the injector-QPC and arc-QPC once both the QPCs are fine-tuned\cite{YKP17,YPM17}. A sharp ZBA peak was observed with the cavity switched off (with $V_2$ = 0 V). On the other hand, the results got modulated significantly with the cavity switched on (with $V_2$ = - 0.90 V). First, a flat ZBA peak was obtained in the low conductance regime of the injector-QPC (G $\leqslant$ 0.5$\times \frac{2e^2}{h}$). Second, additional FBA peaks, occurring around $\pm$0.2 mV, co-existed with the ZBA and thus formed a triple-peak feature when 0.5$\times \frac{2e^2}{h}$ $\leqslant$ G $\leqslant$ 0.8$\times \frac{2e^2}{h}$ [Fig.~\ref{fig:1}(c) is a zoom-in of the ZBA and FBAs]; the triple-peak feature is similar to that reported in QDs\cite{ RFB08} but not yet observed in the QPC. Third, when the system was driven into the high conductance regime (G $\geqslant$ $0.8 \times \frac{2e^2}{h}$), the ZBA evolved into a dip while the FBA remained unchanged [highlighted by the blue trace in Fig.~\ref{fig:1}(b) and (c)], which agrees with the previous results in QPCs\cite{ILK13,BMF14}. On the other hand, switching the cavity on or off in sample D did not result in ZBA nor FBA (see Supplemental Fig.~S5), which was also noticed in a recent work where a flat QPC was coupled to a cavity\cite{SPK17}. 

The observed ZBA only in low conductance regime, the coexistence of ZBA and FBA in the moderate conductance regime, and FBA only in high conductance regime can be understood in terms of evolution of localized electrons within the QPC as shown in Figs.~\ref{fig:2} (the detail of the simulation and further discussion can be found in note 1 and 6 of the Supplemental Material). The enhanced reflection probability at the entrance and exit of the slot-shaped injector-QPC (sample A-C) results in formation of emergent localized electrons (ELS)\cite{MHW02,ILK13} in sample A-C; whilst the smooth varying potential profile in sample D makes it difficult to sustain an ELS, as shown in Fig.~\ref{fig:2}(a) and (b). The electron density evolves from a single peak into multiple peaks on tuning the gate voltage (each peak corresponds to an ELS\cite{MHW02,ILK13}) in sample A-C. It is interesting to note the ELSs eventually merge with the smooth background at high conductance regime (see the trace for 1.2G$_0$) which explains why the ZBA and FBA were absent in the corresponding regime. After showing the general trend of evolution of electron density, we focus on the double-ELS regime whose spin configuration is directly related to FBA. The simulated results shown in Fig.~\ref{fig:2}(c) and (d)  suggest there is a transition from spin triplet state (V$_1$=-1.35 V, the  injector-QPC conductance is $\sim$0.6G$_0$) to spin singlet state (V$_1$=-1.33 V, the conductance is $\sim$0.8G$_0$). It has been shown in previous reports\cite{PRM06,RFB08} that the ZBA should be observed for the triplet state while it will be absent for the singlet state; on the contrary, FBA is allowed for both triplet and singlet states\cite{RFB08}. The experimental observation shown in Fig.~\ref{fig:1}(c) agrees well with the simulation that coexistence of ZBA and FBA occurs in the moderate conductance regime (triplet state) while only FBA is present in high conductance regime (singlet state). 

 \begin{figure}

	\includegraphics[height=3.6in,width=3.0in]{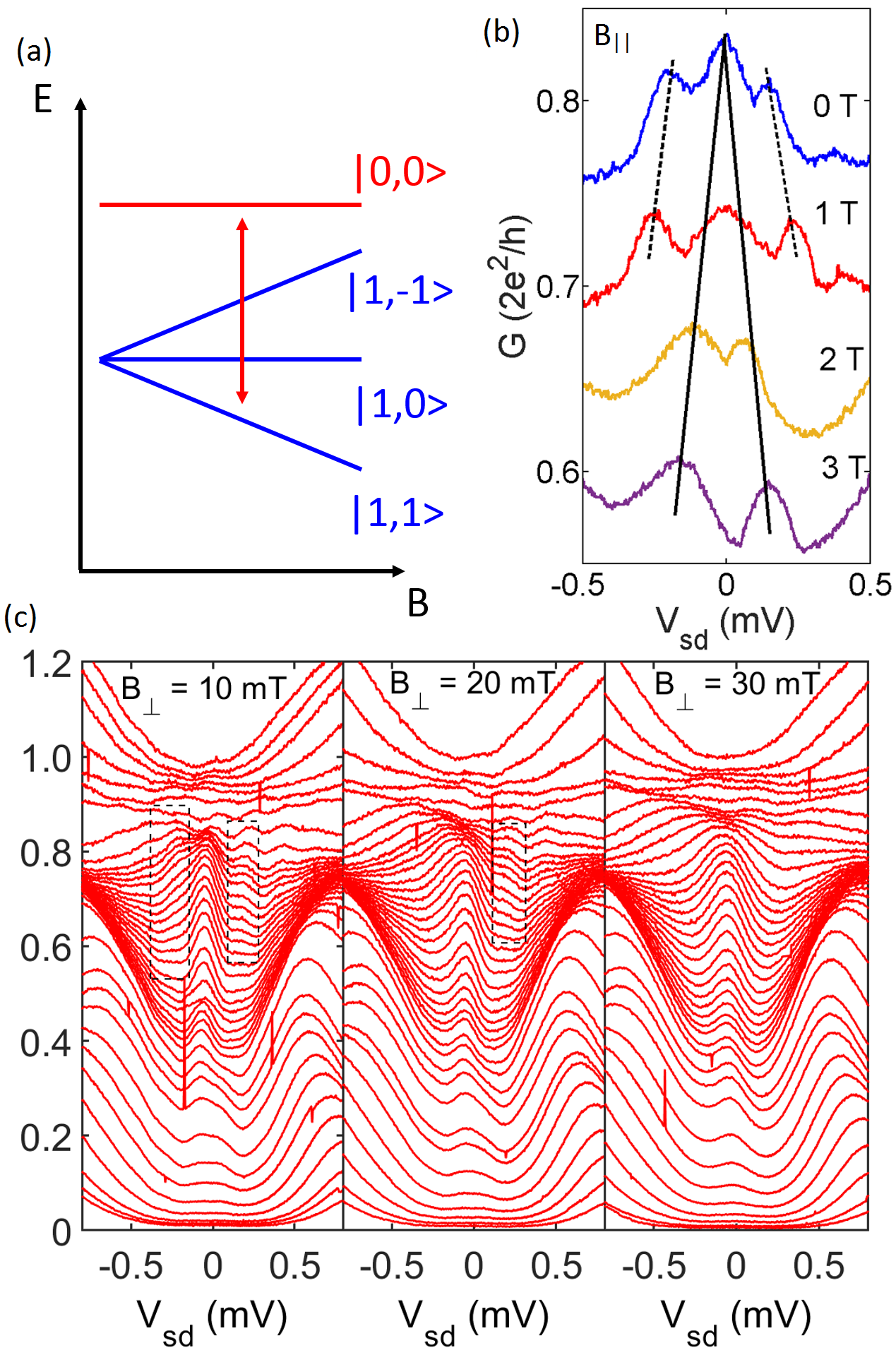} 
	
	\caption{Magnetic field dependence of the ZBA and FBA. (a) Schematic for the evolution of singlet (red) and triplet states (blue) in the presence of in-plane magnetic field in triplet regime. (b) ZBA and FBA with $V_1$ set to -1.35 V at different in-plane magnetic field with the cavity switched on. Data have been offset vertically for clarity. (c) The ZBA persists with the application of the transverse magnetic field while the FBA smears out at 30 mT.
	} 
	\label{fig:3}
\end{figure} 

\subsection{Magnetic field dependence}
To further support our argument, we present results in the presence of a magnetic field. The in-plane magnetic field lifts the degeneracy of the triplet state while it does not affect the singlet state as shown in Fig.~\ref{fig:3}(a). In triplet regime where triplet states correspond to lower energy compared to singlet state at zero magnetic field, the energy difference between the singlet state and the lowest triplet state increases linearly with increasing magnetic field\cite{RFB08,HDF06}. It is seen that with an in-plane magnetic field of 1 T, the ZBA becomes broadened while FBA moves towards larger $V_{sd}$ as expected\cite{RFB08,HDF06}. By increasing the magnetic field further to 2 T or 3 T, the ZBA splits into two. The FBA seems to smear out in the presence of a large magnetic field which is likely due to the transverse magnetic field component induced by the imperfection in field orientation. We show the influence of a transverse magnetic field in Fig.~\ref{fig:3}(c). We noted that at a small transverse magnetic field of 30 mT [the result at 0 T is the same as left panel of Fig.~\ref{fig:1}(b)], the FBAs were smeared out and only the ZBA was left over. The small transverse magnetic field is insufficient to introduce a noticeable Zeeman energy, but enough to influence electron propagation in the cavity\cite{MTM94,YKP17,YPM17}, so that the cavity is not able to refocus the electrons back to the QPC. In other words, the cavity cannot efficiently modulate the effective electron density within the QPC in the presence of transverse magnetic field.

  \begin{figure}

	\includegraphics[height=3.6in,width=3.2in]{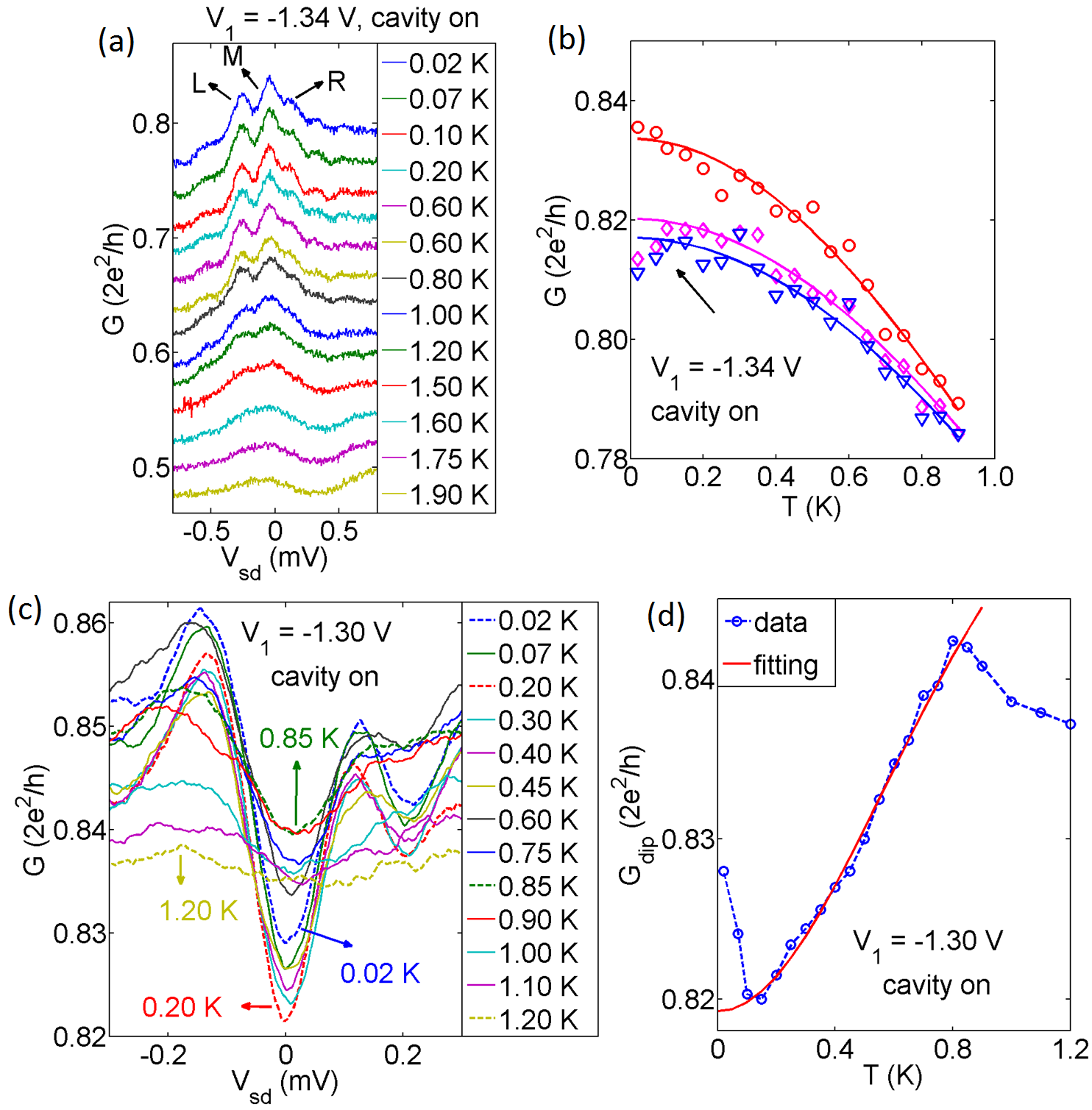}

	\caption{Temperature dependence of the ZBA and FBA. (a) Temperature dependence in the ZBA-FBA coexistence regime. The data have been offset vertically for clarity. (b) Fittings for anomalies referred as M (red round markers; $T_K$ = 1.413 K, $s$ = 0.249), L (magenta diamond markers; $T_K$ = 1.804 K, $s$ = 0.248) and R (triangular blue markers; $T_K$ = 1.812 K, $s$ = 0.247) using Eq.(1), respectively. (c) Temperature dependence in the high conductance regime (V$_1$ = -1.30 V) where only the FBA was observable (without offset). (d) Conductance of the central dip in (c) as a function of temperature; the red solid line is a fitting using Eq.(2), $T^{\ast}$ = 0.72 K and $s$ = 0.22.  }
	\label{fig:4}
\end{figure}

\subsection{Temperature dependence}
The Kondo effect is known for its characteristic temperature dependence, thus it is interesting to investigate the temperature dependence of the multi-impurity Kondo effect as well. Figure~\ref{fig:4}(a) shows the temperature dependence of the ZBA-FBA coexistence regime with cavity switched on (see Supplemental Fig.~S6 for data with the cavity off). Both the ZBA and FBA were attenuated with increasing temperature when the cavity was switched on (V$_2$ = -0.90 V), the left and right FBA smeared out alternatively as the temperature was increased up to 1 K, leaving a broad ZBA at higher temperature.

 \begin{figure}

	\includegraphics[height=4.2in,width=3.6in]{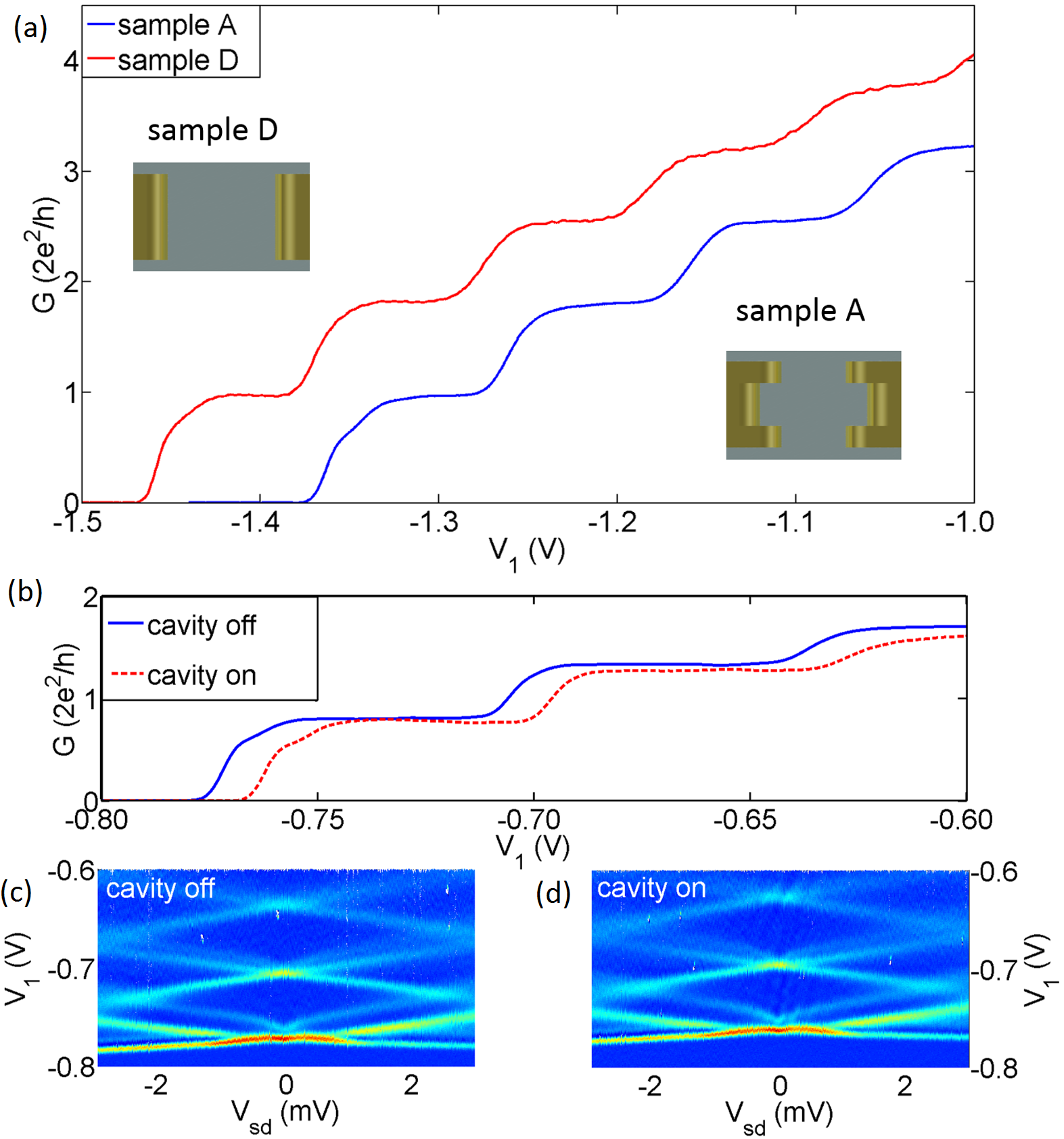}

	\caption{0.7 conductance is not correlated with ZBA nor FBA. (a) Measured conductance of the injector-QPC in sample A (blue trace) and sample D (red trace). The upper and lower inset show schematic of injector-QPC for sample D and A, respectively. (b) Injector conductance measured with cavity switched on (red dotted trace) and off (blue solid trace). (c) and (d) show transconductance $\frac{dG}{dV_1}$ with cavity off and on, respectively.}
	\label{fig:5}
\end{figure}

A more detailed analysis is presented in Fig.~\ref{fig:4}(b) using the standard Kondo function in QPC\cite{CLG02,COK98}, $$G=\frac{2e^2}{h}\{0.5 \times [1+(2^{1/s}-1)(\frac{T}{T_K})^2]^{-s}+0.5\} \eqno(1)$$ where $T_K$ is the Kondo temperature and $s$ is a fitting parameter characterizes the screening between conduction and localized electrons. It is noted that the ZBA agrees well with the theoretical fitting no matter whether the cavity is switched off or on. On the other hand, although the standard Kondo model is not meant for the FBA, it is surprising to note that the standard model can reproduce the temperature dependence of FBA when T $\geqslant$ 0.1 K, which might be due to the fact that the triplet state may decouple into two independent spin-half units at higher temperature\cite{RFB08} and thus the standard Kondo effect dominates. The anomalous suppression of the FBAs in the lowest temperature regime [see the left most data points in Fig.~\ref{fig:4}(b), the black arrow highlights the critical point] diverges significantly from the standard Kondo model. To shed more light on the weakening of FBAs, further experimental results in even lower temperature regime are required.

To make a direct comparison between our observation and results presented in Ref.11, we present the temperature dependence in the FBA-only regime as shown in Fig.~\ref{fig:4}(c) and (d). Apart from the unusual rise of $G_{dip}$ (conductance of the central dip) below 0.1 K [roughly the same value of the critical point in Fig.~\ref{fig:4}(b)], the nonmonotonic trend agrees well with G$_{dip}$ in both QD\cite{RFB08,HDF06} and QPC\cite{ILK13}, and can be fitted by the two-stage Kondo screening model\cite{WSJ02}. In the first stage when the temperature T$\geqslant$0.8 K, the energy difference between the singlet and triplet state E$_S$ - E$_T$ is smaller than k$_B$T, so that the Kondo screening of one of the ELSs instead of the singlet state dominates\cite{RFB08}. The system behaves spin-half-like, therefore the temperature dependence follows the standard Kondo model [i.e. Eq.~(1), the fitting for this section is not shown]. When the temperature T$<$0.8 K, the whole singlet state is screened and the temperature dependence can be described by the re-entrant Kondo formula\cite{RFB08},  $$G=\frac{2e^2}{h} \{1-\alpha \times [1+(2^{1/s}-1)(\frac{T}{T^{\ast}})^2]^{-s}\}+G_c\eqno(2)$$ where $\alpha$ = 1 for QDs\cite{RFB08}, however, we set it as a free fitting parameter to account for difference between the QPCs and QDs, $G_c$  is the background conductance, $s$ is set to 0.22\cite{COK98}, and $T^\ast$ originates from the renormalized singlet binding energy $k_B$$T^\ast$ which is estimated to be 0.72 K. It is noted that $T^\ast$ is in good agreement with the critical temperature ($\sim$0.8 K) below which G$_{dip}$ gradually reduces, Fig.~\ref{fig:4}(d). 

\subsection{Discussion on other possible mechanisms of FBA}

There are several other mechanisms that can possibly result in FBA in addition to singlet-triplet Kondo effect, namely, the spin-orbit interaction\cite{CNF13}, lifting of K-K$^\prime$ degeneracy\cite{SSM15}, coupling the QPC to a high frequency bosonic environment\cite{KSS96}, non-Kondo disorder within the 1D channel\cite{SSD09}, and pinning of the Kondo resonance to the chemical potential owning to the asymmetric device design\cite{SBK99}. The conditions (such as the spin-orbit interaction) for the first three mechanisms are unlikely to be fulfilled in the current experiment setup, whereas the latter two cannot result in the coexistence of the ZBA and FBA. More importantly, these mechanisms predict rather different temperature and magnetic field dependence compared to the one we have observed. Hence, we can exclude the mentioned alternative interpretation for the observed FBA (a detailed discussion can be found in note 7 of the Supplemental Material).

\section{0.7-structure}
Apart from the conductance quantization in 1D system\cite{WTN88,VVB88}, a so called '0.7-structure'\cite{TNS96} (a conductance anomaly that occurs at 0.7$\times \frac{2e^2}{h}$) has been widely observed and attributed to the many-body effect. The origin of 0.7-structure remains a subject of continuous debate. A recent work indicated there is a correlation between the 0.7-structure and the FBA\cite{ILK13,BMF14} and thus suggested the 0.7-structure could be closely associated with the Kondo effect\cite{MHW02}. However, such a correlation is absent in our experiment. In Fig.~\ref{fig:5}(a) we show a comparison between sample A (ZBA is present) and D (ZBA is absent) with the cavity switched off so that only the single-impurity Kondo effect matters. A pronounced 0.7-structure was present in both cases. Figure~\ref{fig:5}(b) shows the result in sample B with the cavity switched on (FBA is observable; source-drain bias spectrum of sample B is present in Supplemental Fig.~S4) and off (FBA is absent). We found that the 0.7-structure was not affected by the presence of FBA (i.e. the multi-impurity Kondo effect). The trend was also clear in the source-drain bias spectrum presented in  Fig.~\ref{fig:5}(c) and (d) with cavity switched off and on, respectively.  Apart from the change in pinch-off voltage, there was hardly any change in the spectrum. Therefore, it seems that the 0.7-structure seen in the present case may not be related to the Kondo effect.     

It has been widely shown in previous works that 0.7-structure can be a more general feature than single-impurity Kondo effect (a recent summary can be found in Ref. 34), so here we suggest the conclusion is also valid in multi-impurity Kondo regime.

\section{Conclusion}
In conclusion, we demonstrated the singlet-triplet Kondo effect in a QPC-cavity hybrid system via the coexistence of ZBA and FBA. The FBA is shown to be highly sensitive to the coupling between the QPC and cavity. The temperature dependence of the FBA uncovers a detailed evolution of the total spin of the localized electrons. The results may open up a different regime of experimentation using the QPCs to explore singlet-triplet effects which so far was largely restricted to QDs.   

This work was supported by the Engineering and Physical Sciences Research Council (EPSRC), U.K.

\end{document}